\documentclass[apj]{emulateapj}
\journalinfo{\textsc{The Astrophysical Journal}}
\slugcomment{Received 2017 November 6; Accepted 2018 March 12}
\usepackage{url}

\usepackage{natbib}
\usepackage{booktabs}
\usepackage{epstopdf}
\usepackage{graphicx}
\usepackage[colorlinks,
        hypertexnames=false,
        pagebackref=false,
        linkcolor=blue,    % color of internal links
    citecolor=blue,        % color of links to bibliography
    filecolor=magenta,     % color of file links
    urlcolor=blue          % color of url
]{hyperref}                %   package that links in pdf
\newcommand{\eg}{e.g.}

\newcommand{\kmps}{\hbox{~\rm{$\mbox{km s}^{-1}$}}}

\newcommand{\Bp}{B_{\rm p}}
\newcommand{\Ef}{E_{\rm f}}
\newcommand{\Efs}{E_{\rm f,s}}
\newcommand{\Ep}{E_{\rm p}}
\newcommand{\Et}{E_{\rm t}}

\newcommand{\Tw}{T_{\rm w}}
\newcommand{\rmd}{{\rm d }}

\shorttitle{Magnetic structure of erupting sigmoid from AR 12371}
\shortauthors{Vemareddy \& Dem\'oulin}
\begin{document}
	\title{Study of Three-Dimensional Magnetic Structure and the Successive Eruptive Nature of Active Region 12371 }
	\author{P.~Vemareddy$^1$ and P. Dem\'oulin$^2$}
	\affil{$^1$Indian Institute of Astrophysics, II Block, Koramangala, Bengaluru-560034, India}
	\affil{$^2$Observatoire de Paris, LESIA, UMR 8109 (CNRS), F-92195 Meudon, France}
	\email{vemareddy@iiap.res.in}
%%%%%%%%%%%%%%%%%%%%%%%%%%%%%%%%%%%%%%%%%%%%%%%%%%%%%
%% Abstract %
%%%%%%%%%%%%%%%%%%%%%%%%%%%%%%%%%%%%%%%%%%%%%%%%%%%%%
\begin{abstract}
We study the magnetic structure of successively erupting sigmoid in active region 12371 by modeling the quasi-static coronal field evolution with non-linear force-free field (NLFFF) equilibria. HMI/SDO vector magnetograms are used as input to the NLFFF model. In all eruption events, the modeled structure resembles the observed pre-eruptive coronal sigmoid and the NLFFF core-field is a combination of double inverse J-shaped and inverse-S field-lines with dips touching the photosphere. Such field-lines are formed by flux-cancellation reconnection of opposite-J field-lines at bald-patch locations.  It implies the formation of a weakly twisted flux-rope from large scale sheared arcade field lines. Later on, this flux-rope undergo coronal tether-cutting reconnection until a CME is triggered. The modeled structure captured these major features of sigmoid-to-arcade-to-sigmoid transformation, that is being recurrent under continuous photospheric flux motions. Calculations of the field-line twist reveal a fractional increase followed by a decrease of the number of pixels having a range of twist. This traces the buildup process of a twisted core-field by slow photospheric motions and the relaxation after eruption, respectively. Our study infers that the large eruptivity of this AR is due to a steep decrease of the background coronal field meeting the torus instability criteria at low height ($\approx 40$ Mm) in contrast to non-eruptive ARs. %In agreement with the previous study, we suggest that the lower/higher height of criticality is related to the magnetic flux normalized helicity flux, derived from photospheric velocity and magnetic fields, accounting the more/less twisted nature of magnetic flux.  

\end{abstract} 
%%%%%%%%%%%%%%%%%%%%%%%%%%%%%%%%%%%%%%%%%%%%%%%%%%%%%%%%%
%% 1. Introduction %%%%%%%%%%%%%%%%%%%%%%%%%%%%%%%%%%%%%%%%%%%%%%%%%%%%%%%%%%
\section{Introduction}
\label{Intro}
% {\S}{\bf --- Generality Description } \\

It is now accepted that the source of energy for all solar activity is from magnetic fields. Active regions (ARs) are higher concentrations of magnetic field regions often associated with violent activity like jets, flares, coronal mass ejections (CMEs) etc. In X-rays and EUV observations, these regions are sometimes seen with a sigmoid structure \citep{rust1996,gibson2002} situated over the central polarity inversion line (PIL). The sigmoid structure is either forward or inverse S-shaped.  Several CMEs are seen to be launched from these sigmoidal ARs and hence sigmoid is one of the most important pre-cursor structure for solar eruptions \citep{manoharan1996,hudson1998,canfield1999,canfield2007}.  
Two types of sigmoids were reported: transient and long lived sigmoids. Transient sigmoids are sharp and bright and they usually become clearly noticeable only for a short time before eruption, while long-lived ones appear more diffuse and can survive for several hours, even days \citep{pevtsov2002,mckenzie2008}. 

%{\S}{\bf --- Different sigmoid models } \\
The magnetic structure of sigmoids was described as sheared arcade and flux rope (FR) topology \citep{moore2001}. In the sheared arcade model, the two magnetic elbows are sheared field lines which are located nearby on both sides of the PIL in the central part of the configuration.
In the FR scenario, a magnetic FR is embedded in a stabilizing potential envelope field \citep{moore1992,hood1979,titov1999}. 
How these sigmoids turn to be eruptive? 
Magnetic reconnection (resistive instability) plays the prime role in the sheared arcade model as opposed to ideal MHD instability in the FR model in triggering the eruption \citep{moore2001,antiochos1999,amari2003a,torok2005,gibson2006}.
The interface separating a coronal FR from its ambient field usually forms a sigmoidal shape when observed from above and FRs are naturally invoked in different models of sigmoids \citep{titov1999,gibson2006,bobra2008,savcheva2009}. Sigmoids are considered to result from enhanced current dissipation in thin sheets which accumulate hot plasma along corresponding shaped field lines \citep{gibson2002,janvier2017} .

%{\S}{\bf --- Electric currents } \\
During the evolution of an AR, magnetic energy is built up (during several hours or even days) in the corona due to flux emergence and photospheric motions \citep{schrijver2009}. The energy is stored in a non-potential field associated to large scale electric currents \citep{priest2002,aulanier2010}.

%{\S}{\bf --- Topology } \\
Topological analysis of magnetic structures constructed from analytic configurations predicts that the current sheets form along magnetic interface layers called separatrices, as well along their generalization called quasi-separatrix layers \citep{priest1995,priest2000,Aulanier2005}. The connectivity of field lines is discontinuous for separatrices, while it changes drastically along QSLs \citep{demoulin1996}.  In an analytical force-free FR model embedded in an arcade field, \citet{titov1999} showed that in the process of the FR emerging rigidly into the corona, a separatrix surface, touching the photosphere at a so called bald-patch (BP, $\mathbf{B}\cdot\nabla B_z|_{PIL}>0$), is present along the PIL section during the earlier phase of emergence.  The BP separatrix surface has a generic S-shape when viewed from top, similar to sigmoid shape. In the later phase of emergence, the S-shaped BP separatrix is transformed into a double J-shaped QSL. The topological analysis of the FR eruptions found co-spatial flare ribbons with hook-shaped QSLs \citep{demoulin1996b,williamsdr2005,janvier2014,zhaoj2016}, leading to extension of 2D version of standard model to 3D \citep{janvier2015}.

%  {\S}{\bf --- Formation of sgmoids } \\
In decaying ARs, the sigmoids were reported to be formed by flux cancellation induced by converging motions of magnetic elements towards PIL \citep{green2009,tripathi2009,vemareddy2015b}.  Helical field lines are formed by reconnection of the sheared field lines \citep{pneuman1983,ballegooijen1989}. Indeed, numerical simulations had shown the FR formation under cancelling magnetic flux scenario \citep[e.g., ][]{amari2003a}. On the other hand, numerical models producing FRs also relied on magnetic flux emergence of twisted FRs originating from the convection zone and rising through the photosphere \citep{magara2001,fany2004,gibson2004,archontis2009,hood2012}. In these models, as the FR emerges a filamentary current sheet forms, reconnection occurs and a new coronal FR is formed. When integrated along the local vertical, the current sheet appears as sigmoid. 

%  {\S}{\bf --- Extrapolations } \\
Observationally studying the 3D magnetic structure of AR is presently not possible.  Therefore, one relies on models based on field measurements at the photosphere as the lower boundary condition. A non-linear force-free field (NLFFF) model is typically used to reconstruct pre-eruptive coronal field \citep{valori2005,savcheva2009,
wiegelmann2008, derosa2009,jiangc2012a}. This is justified by the low--$\beta$ coronal condition. 
However, this is not realised at the photospheric level and a pre-processing procedure is applied to drive the lower boundary towards force-free condition \citep{wiegelmann2006}. Moreover, the photospheric flow speed is very slow, typically well below 1 \kmps \citep{vemareddy2012b} , compared to the coronal field relaxing speed to equilibrium, which is several hundred \kmps . Therefore, the coronal field evolution, driven by slow photospheric motion, can be approximated as a quasi-static evolution and it is modelled by time sequence of successive static force-free equilibria. It enables to study the build-up of pre-eruptive 3D structure like sigmoid, then to find hints on the most appropriate configurations leading to eruptions (e.g., \citealt{sunx2012,sunx2013b,savcheva2012a,jiangc2014,vemareddy2014a,vemareddy2016b}).

%  {\S}{\bf ---  Aims of this study} \\
In present article, we construct 3D magnetic structure of sigmoidal AR by NLFFF model and study the pre-eruptive configuration of observed CMEs. For tis purpose, we employ vector magnetic field measurements at the photospheric level obtained from \textit{Helioseismic and Magnetic Imager} (HMI; \citealt{schou2012}) on board \textit{Solar Dynamic Observatory} (SDO; \citealt{pesnell2012}). We study below the AR 12371 which was producing successive fast CMEs during its disk transit. In EUV observations, the pre-eruptive configuration of all CMEs is a sigmoid. Being CME productive, we study how the sigmoid developed repeatedly and how it was brought eventually to instability.  This case 
is in contrast with most previous studies of decaying ARs where only one single sigmoid eruption was present during the observation window \citep{demoulin2002,green2009,green2011,vemareddy2015b}.  Based on the time sequence of this static NLFFF extrapolations, we focus on studying 
  1) the sigmoid build up in the pre-eruptive configuration, 
  2) the build up of the coronal electric currents and their dissipation during eruption, 
  3) the magnetic energy,
  4) the evolution of field line twist distribution and, 
  5) the role of the background field in driving the eruptions. 
We outline the observations during CME onset in Section~\ref{outline}. Results of the above studies are described in Section~\ref{res}. Summary with a discussion of the results is given in Section~\ref{summ}.  
\begin{figure}[!ht]
	\centering
	\includegraphics[width=.49\textwidth,clip=]{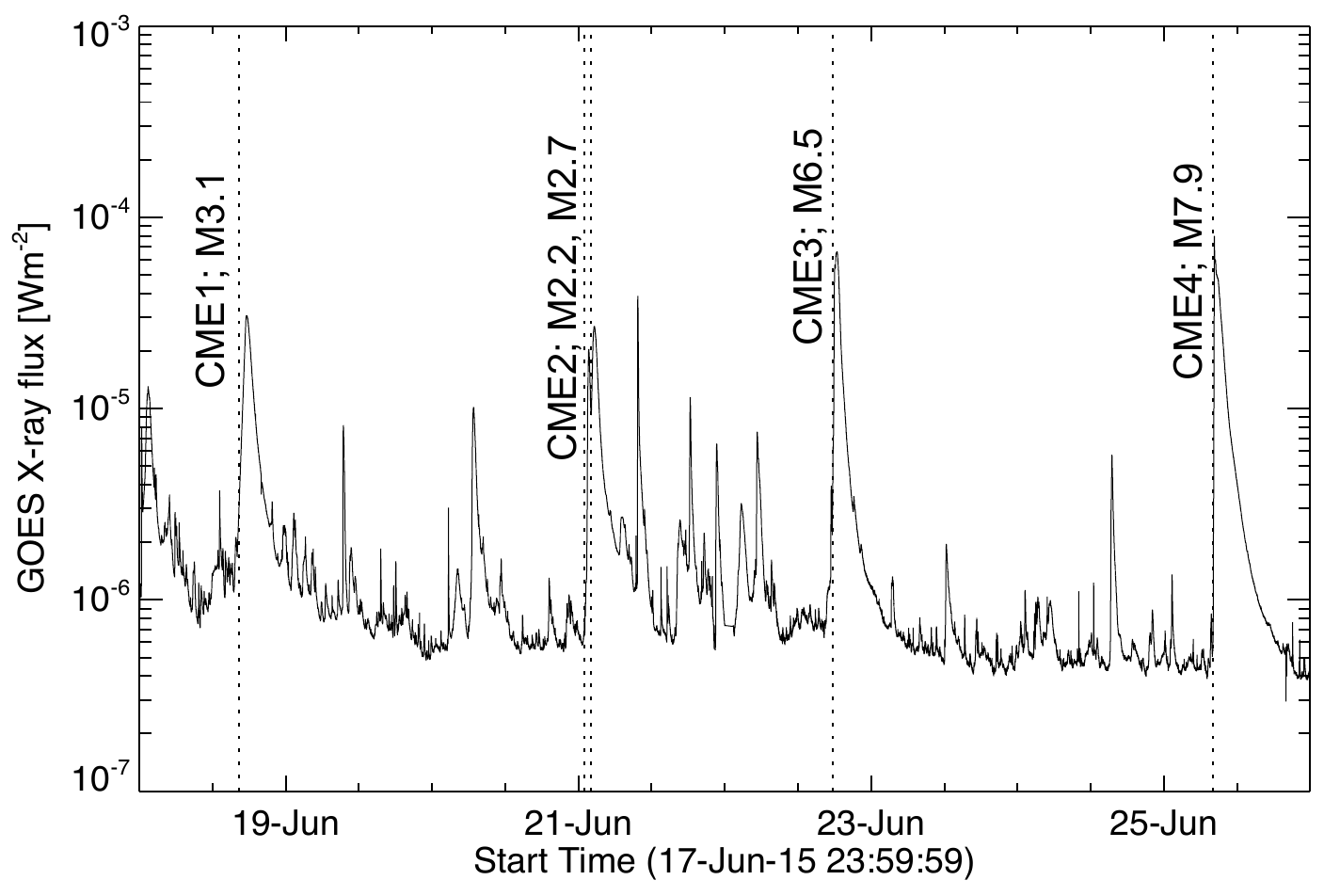}
	\caption{Disk integrated GOES X-ray flux (1-8\AA) during June 18-25, 2015. Vertical dotted lines mark initial times of CME-associated M-class flares from AR 12371.	}
	\label{Fig_goes_plot}
\end{figure}

\begin{figure*}[!ht]
	\centering
	\includegraphics[width=.92\textwidth,clip=]{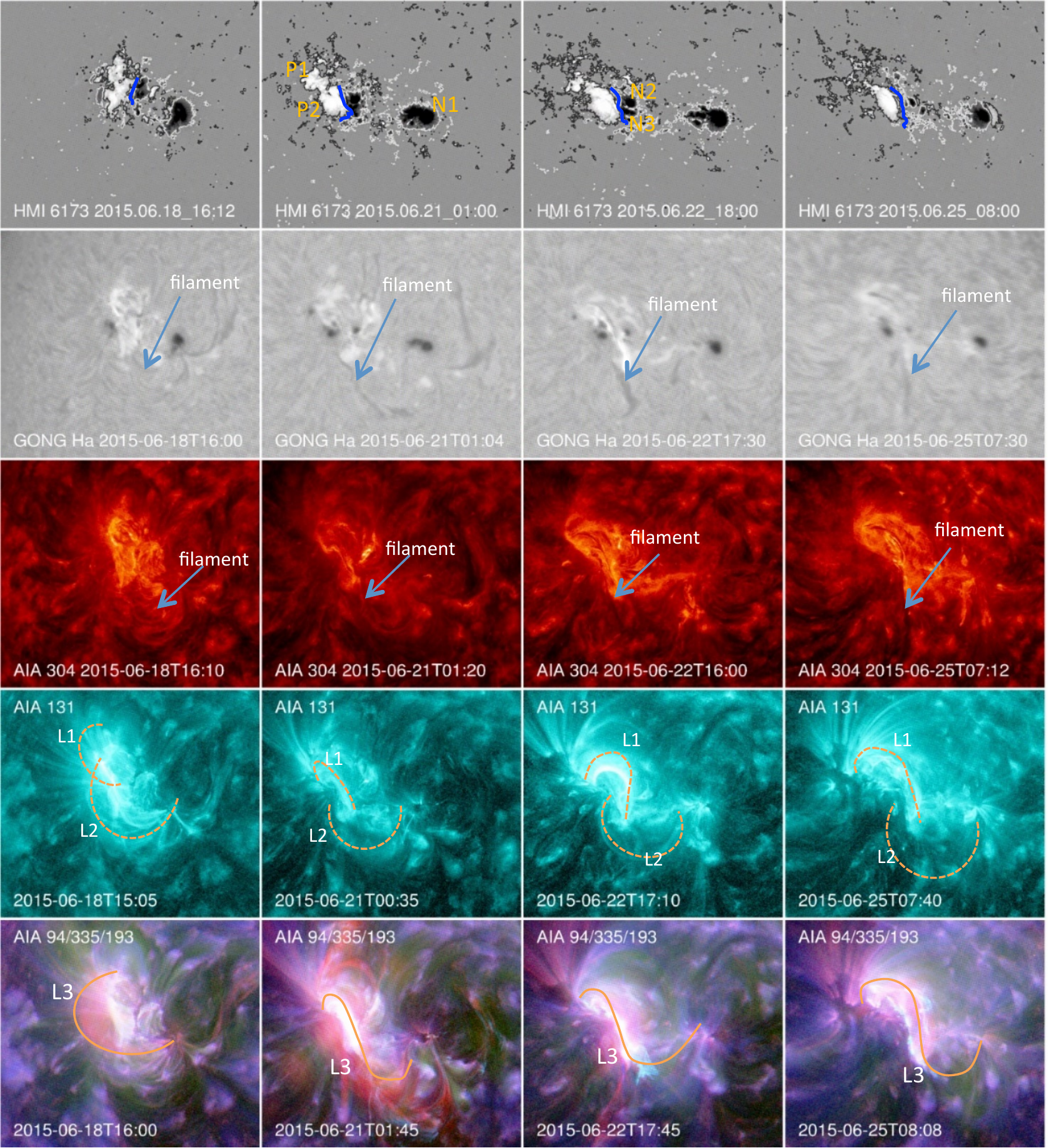}
	\caption{\footnotesize{Multi-wavelength observations of the AR 12371 during the initiation of four CMEs shown in the four columns. Each panel of the same event have the same field of view. 
			\textit{First row:} HMI line-of-sight magnetic field maps with labeled polarity regions. Blue curve is the main PIL. 
			\textit{Second row:} GONG H$\alpha$ observations showing traces of chromospheric filaments and sunspots (related to HMI magnetic polarities).
			\textit{Third row:} AIA 304 \AA~observations showing the presence of filament (pointed by blue arrow) before the onset of CME eruption. 
			\textit{Fourth row:} AIA 131 \AA~observations before eruptions.  In particular they show the coronal loop sets  as magnetic elbows and are traced with dashed curves L1 and L2 by inspecting time series of images. In these observations, the pre-cursor brightening related to slow tether-cutting reconnection is seen with L1.  
			\textit{Fifth row:} Composite images of coronal observations in AIA 94~\AA\ (red), 335~\AA\ (green), and 193~\AA\ (blue) passbands.  The loop set L3 (outlined with an orange curve) is understood as the dynamic reconnection product of L1 and L2. See Section~\ref{outline} for more details. }
	}
	\label{Fig2}
\end{figure*}
\section{Outline of Observations and Context of the Study}
\label{outline}
%  {\S}{\bf --- Global Descript         ion } \\
AR 12371 is an AR formed before its appearance at the eastern limb. It produced successive fast CMEs in a span of its disk transit (18-25 June, 2015). Recent study of this AR by \citet{vemareddy2017b} presented a detailed analysis of the magnetic evolution in relation with the initiating mechanism of the  successive  CMEs. Here we give a brief outline of earlier results in the context of this study. Observations at the Sun register the initiation times of four eruptions as 15:05\,UT on 18 (SOL2015-06-18T15:05, CME1), 00:45\,UT on 21 (SOL2015-06-21T00:45, CME2), 16:15\,UT on 22 (SOL2015-06-22T16:15, CME3), and 07:30\,UT on 25 (SOL2015-06-25T07:30, CME4) of June 2015, respectively, by the commencement of enhancing EUV brightening associated to fast structural changes. Subsequently, the disk integrated GOES X-ray light curve (Figure~\ref{Fig_goes_plot}) delineates that the CMEs are associated with long duration M-class flares starting from 18/16:25\,UT, 21T02:00\,UT, 22T17:39\,UT, 25T08:02\,UT respectively.  These CMEs emerged the C2 field-of-view at 18/17:24\,UT, 21T02:36\,UT, 22T18:36\,UT, 25T08:00\,UT, respectively. All four CMEs are fast ($>1000$\kmps) in LASCO field-of-view.

 % {\S}{\bf --- B evolution} \\ 
In Figure~\ref{Fig2}, we show simultaneous multi-wavelength observations, taken from different instruments, of the AR during the onset phase of CMEs. Different wavelength snapshots at different times are selected to best show the observed structures. All image panels are aligned accounting spatial scales. In the first row, we plot line of sight magnetograms obtained from HMI. The AR successive passage reveals presence of leading negative flux (N1) with a following bipole with negative (N2, N3) and positive (P1, P2) flux regions. The inner bipole (N2, N3) is seen with large shear and converging motions (see Figure 6 of \citealt{vemareddy2017b}) with respect to (P1, P2), forming sheared polarity inversion line (PIL) indicated with a blue curve. The field distribution of (N2, N3) becomes diffused and disintegrated in successive days while the leader polarity is increasingly separated from the following polarity. 

% {\S}{\bf --- Emissions } \\
In the second row of Figure~\ref{Fig2}, we plot GONG-H$\alpha$ images. They show the presence of a dark filament between leading and following sunspots before the eruptions. In tandem, 304 \AA~observations from the Atmospheric Imaging Assembly (AIA, \citealt{lemen2012}) are well in support with H$_{\alpha}$, showing co-spatial existence of continuous filament channel. A filament is present on the PIL portion southward of the AR. It could only be related to the faint magnetic field observed southward of the AR and be irrelevant to the eruption, since no apparent change of the filament can be recognized after the commencement of EUV brightening from the time sequences associated to Figure~\ref{Fig2}. The  131\AA~channel of AIA (fourth row panels) evidences sheared loop morphology comparable to those previously reported in other ARs \citep[\eg ][]{moore2001}.  They consist of two opposite J-shaped loops outlined by L1 and L2 (dashed orange curves) identified from sequence of images. We call them magnetic elbows.   The inner ends of these loops are nearby and crossed each other with a sharp interface low over the PIL. Finally, the composite images prepared from the hot AIA channels are shown in the bottom row of Figure~\ref{Fig2}. They provide the evidence for continuous sigmoidal structure (labeled L3) during the initiation of all four CMEs. 

%{\S}{\bf --- Interpretation of the observed loops} \\
The configuration outlined by L1 and L2, especially in the three last events, is typically found in models having a sheared core \citep[\eg ][]{antiochos1994,aulanier2010}.  In the tether-cutting scenario, magnetic reconnection is initiated at the interface of L1 and L2 closest legs when converging motions push the legs against each other. The large-scale reconnected field lines have a shape comparable to L3.  This reconnection progressively transforms sheared field lines, such as observed with L1 and L2, to twisted field lines, as outlined with L3. As reconnection proceeds further, the FR bulges and rises in height.  In numerical simulations \citep{aulanier2010}, once the FR is at a too large height, the configuration becomes unstable and the FR escapes the coronal closed field environment into the outer corona as a CME. The AIA composite images (fifth row panels in Figure~\ref{Fig2}) evidence this flux-rope (labelled as L3) during initiation of CMEs. 

\begin{figure*}
	\centering
	\includegraphics[width=0.99\textwidth,clip=]{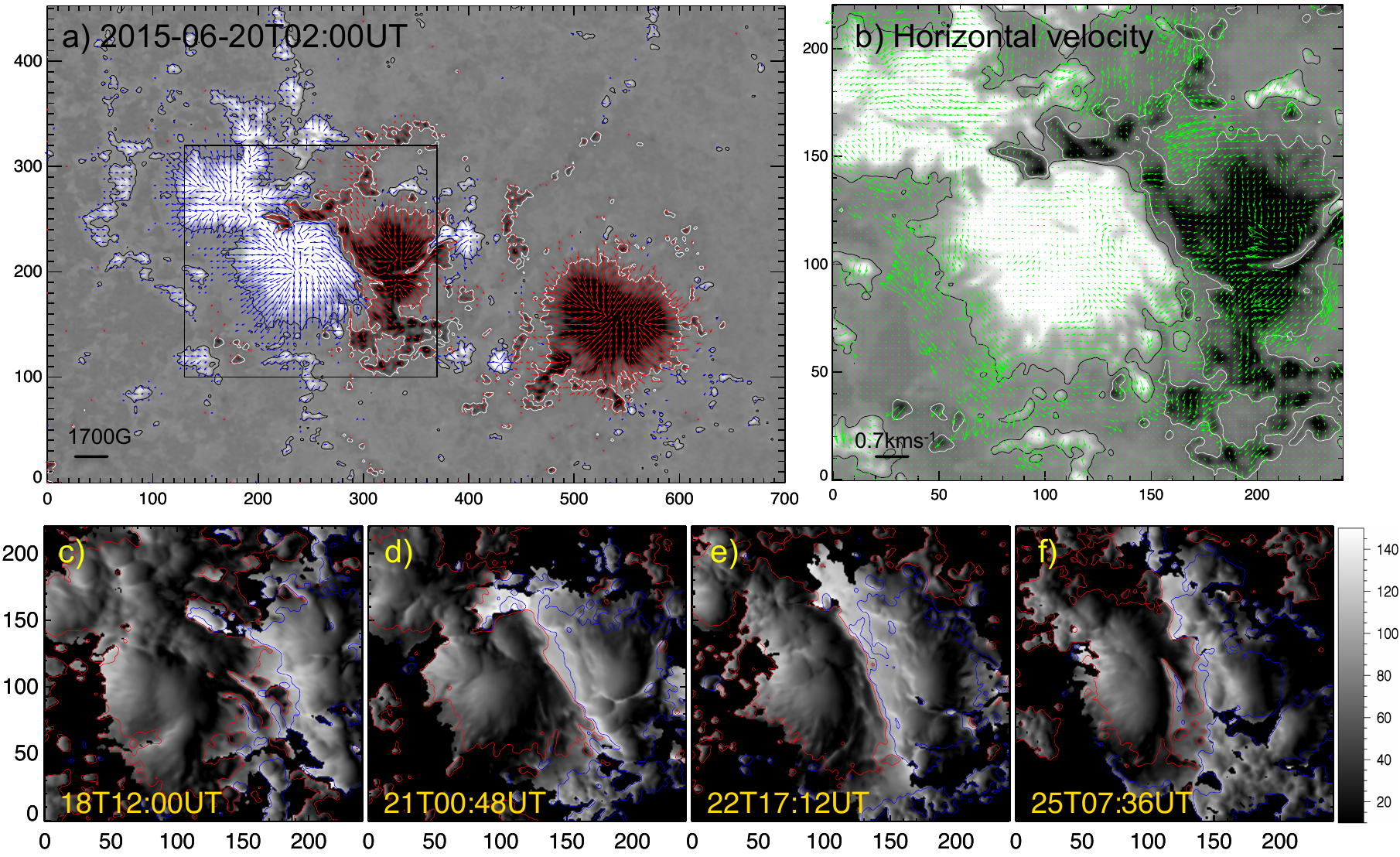}
	\caption{a) HMI Vector magnetogram of AR 12371 on June 20. Blue/red arrows represent horizontal field. Background is vertical field image with contours at $\pm$150\,G. b) Tracked horizontal velocity field on vertical field map. Field-of-view covers the inner bipolar region as shown with rectangle box in panel a). Careful inspection implies a slow shear and converging motion between the opposite polarity regions. 
		c-f) Maps of shear angle in the inner bipolar region covered by rectangle box in a) (see the side color bar). Contours (red/blue) of $B_z$ at $\pm$150 are overlaid. In the vicinity of the PIL, both the positive and negative polarities have sheared locations greater than 45$^\circ$. Finally, as a result of flux cancellation along the PIL and dispersion of the photospheric field, the polarity regions decayed in area from June 18 to 25. In all panels, the axis units are in pixels of 0.5 arcsec. }
	\label{Fig3}
\end{figure*}

% {\S}{\bf --- Summary of results in Vemareddy 2017}  \\
The study of magnetic field evolution (please refer to Figure 7 in \citet{vemareddy2017b}) implies decreasing magnetic flux in both AR polarities, which is an indication of magnetic flux cancellation. Velocity field of flux motions derived from DAVE4VM \citep{schuck2008} show the shear and converging motions, which are in support of above tether cutting reconnection, within the inner bipolar regions (Figure~\ref{Fig3}). These motions were suggested to be prime factors to the flux cancellation and the net flux decrease \citep{chae2002,chae2004}. Under this flux evolution, the net current and $\alpha_{av}$ (mean twist parameter of magnetic field) show increasing trend till June 22. This is an indication of build up of non-potentiality \citep{martinsf1998,wangmuglach2007}.  Further, with these observational findings of magnetic evolution, \citet{vemareddy2017b} described that the repeated sigmoid formation and its eruption by successive CMEs is a result of energy storage and release by persistent shearing and converging motions. With monotonous injection of negative helicity, the corona over-accumulates helicity and expels excess helicity by CME. Predominant influx of helicity of either sign over time is the key for the formation of coronal sigmoid or FR, which would be initiated to ejection by ideal or non-ideal instability \citep{moore2001,torok2005}.  Using vector magnetic field observations at the photosphere, in the present article, we study the different aspects in magnetic structure of the pre-eruptive sigmoid with NLFFF models.        

\begin{figure*}[!ht]
	\centering
	\includegraphics[width=.98\textwidth,clip=]{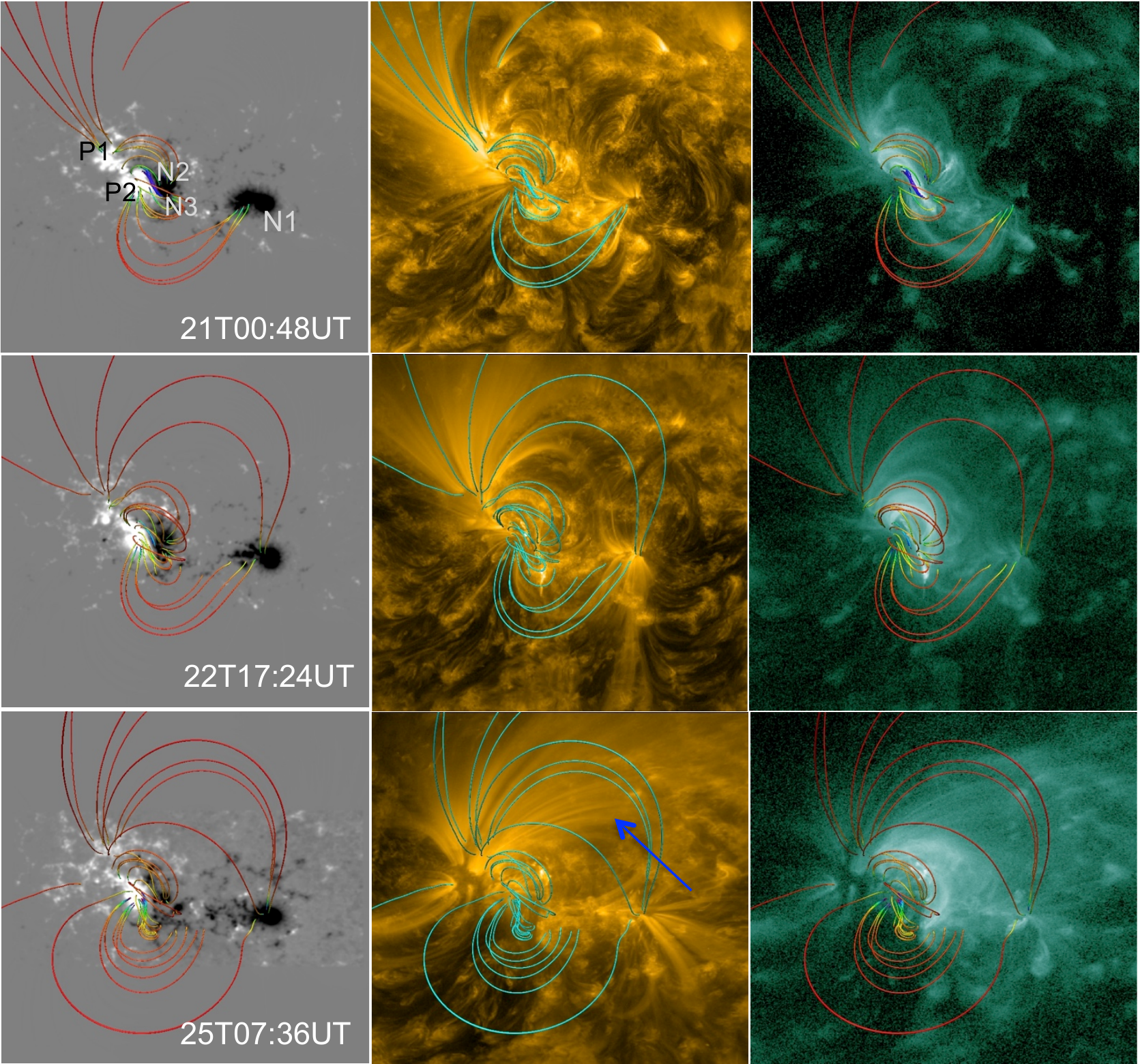}
	\caption{ \footnotesize{
			NLFFF magnetic structure of AR 12371 during the three eruption events.
			\textit{First column:} Field lines plotted on top of the photospheric magnetogram ($B_z$) with labeled polarity regions. The global structure is a sigmoid with two opposite J-shaped sections and an inner sheared core. 
			\textit{Second column:} Same field lines on AIA 171 \AA~ images. Color scale is chosen in contrast to background images. 
			\textit{Third column:} Same field lines on AIA 94 \AA~images. Hot plasma emission is mainly co-spatial with the strongly sheared core. Note that for event 4 (bottom rows) the projection effects tend to deviate the model structure, especially long field lines from the observed 171\AA~loops indicated by arrow. Field lines are color coded (blue (red): $\sim$1200 (2)G) with the horizontal field strength in height in column 1 and 3. }
	}
	\label{Fig4}
\end{figure*}
\section{Results}
\label{res}
\subsection{Non-linear Force-free Modelling of the AR Field}
\label{NLFFF}
%  {\S}{\bf --- Set up } \\
The AR magnetic structure is reconstructed by performing nonlinear force-free field (NLFFF) extrapolation of the observed photospheric vector magnetic field \citep{bobra2014,hoeksema2014}. The procedure minimizes a volume integral including a linear combination of the Lorentz force and of the magnetic field divergence \citep{wiegelmann2004,wiegelmann2010}. The field of view of the boundary field covers the full AR such that flux is nearly balanced over the entire time interval. During all four CME events, the flux imbalance is less than 8\%.  To further satisfy the force-free conditions, the magnetic field components are subjected to a pre-processing procedure \citep{wiegelmann2006}. Next, we rebinned the observations to 1 arc-second per pixel. In order to weaken the effects of the lateral boundaries, the observed boundary is inserted in an extended field of view and computations are performed on a uniformly spaced computational grid of $512 \times 512 \times 256$ representing physical dimensions of $373 \times 373 \times 186$ Mm$^3$. To reduce the effect of pseudo top and side boundaries on the evolving interior field, a boundary wall (of width 64 grid points) of cosine weighting function is used in the function which is minimized.

%  {\S}{\bf --- Pratical extrapolation } \\
The NLFFF code is initiated with a 3D potential field constructed from the vertical field component of the observed field \citep{gary1989}. After introducing the photospheric vector magnetic field at the lower boundary, the magnetic field is relaxed the most possible to decrease both the Lorentz force and the divergence of the magnetic field.  Typically, the observed photospheric field, even after processing, is not compatible with the NLFFF state, but it can be relaxed to a state with an average angle typically around 9 degrees between the electric current and magnetic field vectors in the computational volume.

%  {\S}{\bf --- vector magnetograms } \\
A typical vector magnetogram used in the NLFFF reconstruction is shown in Figure~\ref{Fig3}a. The rectangular box covers the inner bipolar region. Horizontal velocity field, derived from DAVE4VM method, indicate a large scale converging and shear motions of opposite polarities in the inner bipolar region as displayed in panel b). Under these motions, the field vectors are sheared where the angle between potential transverse field and the observed transverse field is non zero.  We measured the shear angle defined as the angle between the potential transverse field and the observed transverse field. In panels c-f), these shear angle maps are displayed at different epochs of evolution. Both the positive and negative polarities have sheared locations greater than 45$^\circ$. Especially, the negative polarity region is more sheared than positive polarity. The magnetic shear is in particular important around the PIL in between (N2, N3) and (P1, P2) (See Figure~\ref{Fig2} for polarity identification).  Tracked magnetograms infer a southward motion of the negative polarity with respect to the positive polarity. This is the main origin of this shear region \citep{vemareddy2017b}. 

%  {\S}{\bf --- Sheared and twisted field lines } \\
As pointed by \citet{moore2001}, two likely factors for the eruption are the flux content of the sheared core field relative to the envelope field and the height at which the reconnection begins. The greater the reconnection height, the smaller the fraction of the core field that can be released as the eruption proceeds. We compute the shear angle in the inner bipolar region  (rectangular box in Figure~\ref{Fig3}a) for the vector magnetograms recorded just before all three eruptions. In this bipolar region, we found that 20\%, 30\%, 30\% of pixels have greater than $45^\circ$ shear angle, while the pixels with shear angle greater than $25^\circ$ are 58\%, 45\%, 48\% respectively for the last three events. Alternatively, the numbers meet the Moore et al's first criteria of excess flux content of sheared core field with shear angle above $25^\circ$. 

In Figure~\ref{Fig4}, we show the NLFFF magnetic structure just before the CMEs. Being located at E50$^\circ$, the CME1 is excluded for further analysis due to severe projection effects. The field lines are selected according to the total current density ($|\mathbf{J}|=\frac{1}{\mu_0} (\nabla \times \mathbf{B})$) and the horizontal field component ($B_h=\sqrt{B_x^2+B_y^2}$), i.e., the seed points are biased with large $|\mathbf{J}|$, $B_h$ in a subvolume covering the inner bipolar region. This will ensure to select field lines that outline most of the observed coronal loops (Figure~\ref{Fig2}). A color scheme according to $Bh$ strength is applied for the field lines. The bottom plane is $B_z$ in first column panels. Field lines connecting the peripheral regions of P1 and N2 form one J-section loop set, and those connecting N1 and P2 form another J-section loop set. These two bundles of J-shaped field lines are formed by highly sheared arcades.
\begin{figure*}[!ht]
	\centering
	\includegraphics[width=.99\textwidth,clip=]{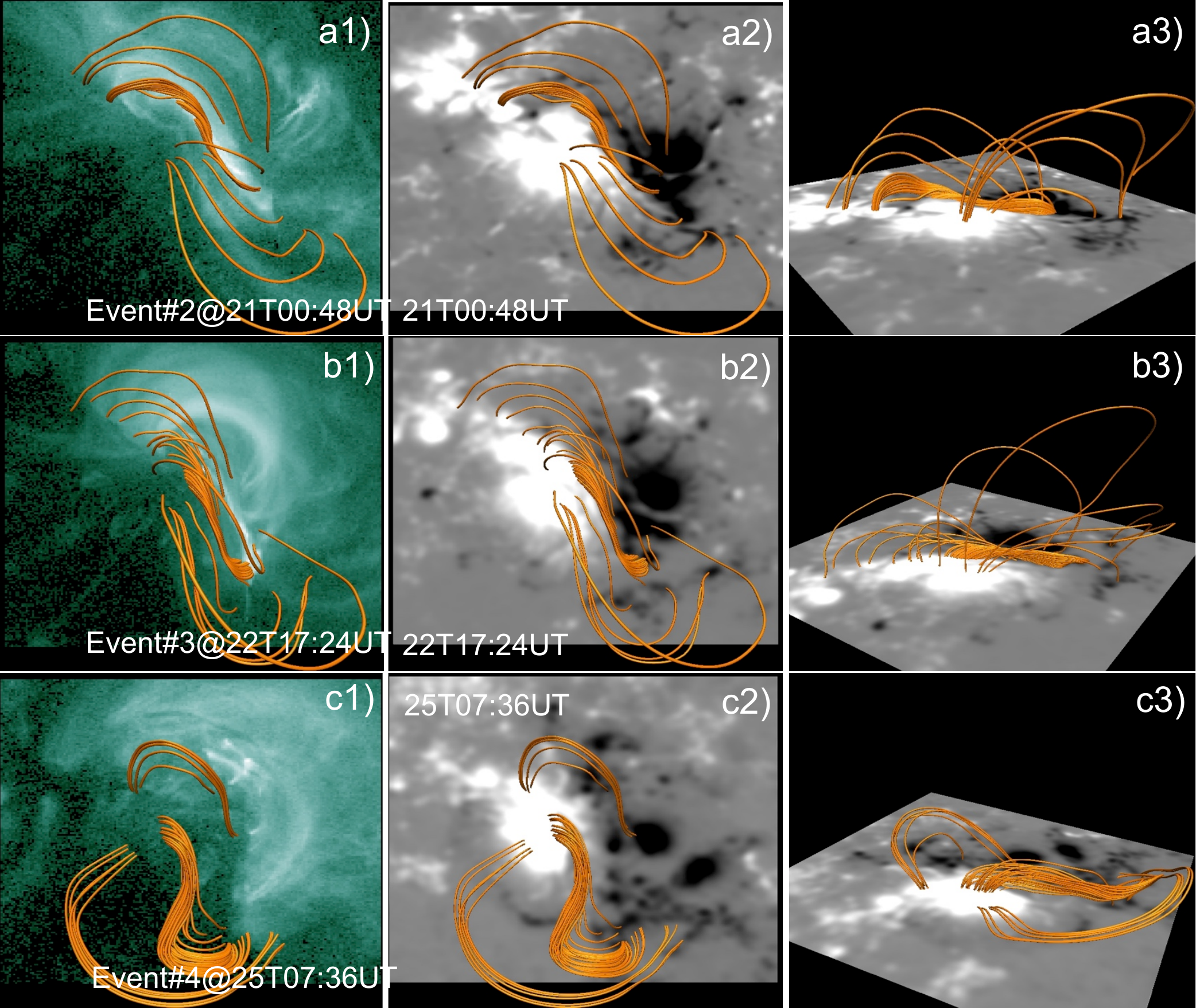}
	\caption{ Field line shapes in the core and EUV observations (organized by rows).  \textit{First column:} Field lines on EUV-94 images, showing twisted field embedded in large-scale sheared arcade. The twisted field 
		is co-spatial with EUV emission, whereas emission from the northern elbow comes from high lying potential field. (see also Figure~\ref{Fig4}) 
		\textit{Second column:} Top view of the core field line structure.  The photospheric $B_z$ distribution observed at the same times is shown with gray levels in the background for all panels.  
		\textit{Third column:} Perspective view of the same field lines showing the coronal extent of the sheared arcade  }
	\label{Fig_core}
\end{figure*}

Owing to line-tied shearing motions of the foot points, the field lines near the PIL, in the AR center, are strongly sheared.  They are enveloping field lines over the middle of the sigmoid. This sheared magnetic configuration, as shown in Figure~\ref{Fig2}, being further driven by converging motions, of two opposite J-sections is expected to favor the formation of FR by tether-cutting reconnection \citep{titov1999,jiangc2012b,savcheva2012a}.  

%  {\S}{\bf --- Comparison to observations } \\
To compare this modeled magnetic structure to coronal observations, the same field lines are over plotted on EUV observations of corona in 171\AA~and 94 \AA~passbands (second and third column panels). The EUV loops are in global agreement with the NLFFF twisted structure in each panel (in agreement with previous results, \citealt{sunx2012,vemareddy2014a}). Remarkably, the intense hot emission is mostly co-spatial with the NLFFF sheared core. The sheared core is less compact for the last event on June 25 compared to others. This indicates a relaxed configuration after a sequence of eruptions and quasi-static evolution in between two events. 

We point that the projection effects in the corona are greater than for the photosphere. Notedly, being away from disk center, projection effects introduce some departures with large scale loops especially for 25 June  (located at 40$^\circ$ west).  However, the low lying core structure is nearly reproduced because disk transformation from local to observer frames mostly take into account these projection effects.  
%  {\S}{\bf --- Results on the core } \\
 
In a study of sigmoid eruption from another AR (AR 11283) \citet{jiangc2014} reported the FR formation by reconnection which was attached to photosphere at the bald patch (BP) until eruption. Following their analysis, we further discuss the eruption cases in AR 12371 as follows. In Figure~\ref{Fig_core}, we examine the core field topology of the NLFFF structure. Field lines anchoring away from the PIL correspond to two J-sections with their arms farther apart from the PIL. For the events on June 21 and 22, the field lines anchored near the PIL are inverse S-shaped and graze the PIL.  Indeed, we note little difference in the observed EUV morphology with the core field. The northern sigmoid elbow in all the cases could be related to high lying potential-like arcade. The southern elbow seem to constitute by a continuity of field lines belonging to both low lying core field and high lying arcade to the leading polarity. 

\begin{figure*}[!ht]
	\centering
	\includegraphics[width=.88\textwidth,clip=]{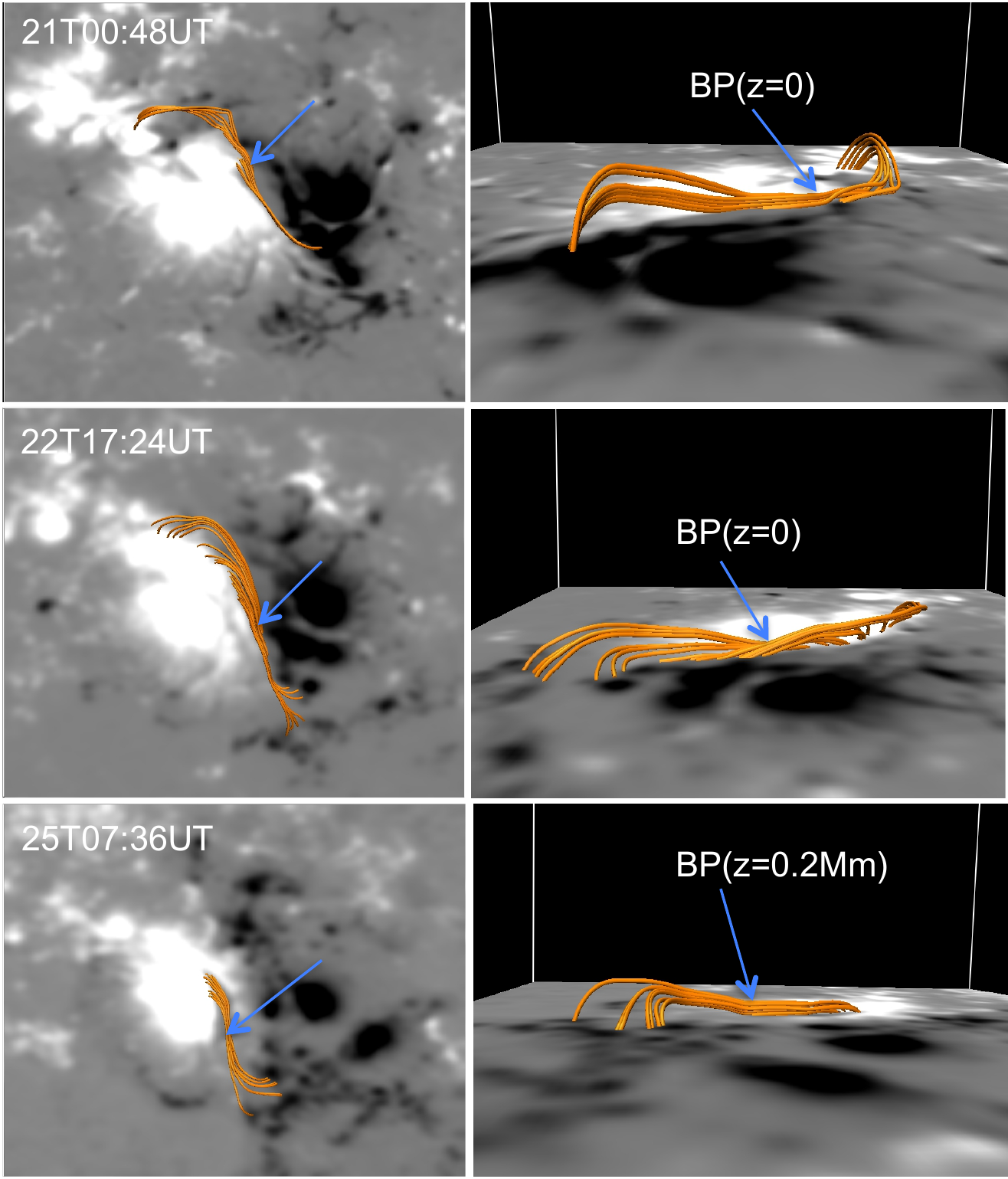}
	\caption{ NLFFF structure in the inner core evidencing the BP separatrix topology. Left column: Top view of field lines plotted on $B_z$ maps. These field lines are combination of double inverse-J and S shape along the PIL. Right column: Perspective view of the same field lines having prominent dips touching tangentially the photosphere. Arrows point to the field line dips near to the surface ($z=0$). This kind of topology is suited for flux cancellation reconnection of inverse J-shape field lines to form inverse-S shape field lines and the FR.}
	\label{Fig_bp}
\end{figure*}

% {\S}{\bf --- General physical expected evolution compared with present event results } \\
BP separatrix is a common topological feature in sheared bipolar regions, where the buildup of FR begins with the appearance of BP \citep{titov1999,aulanier2010}. In Figure~\ref{Fig_bp}, the field lines along the PIL are displayed in a close view for BP topology. The structure for the event on June 21 clearly show BP separatrix field lines. Field lines surrounding the BP separatrix are combination of inverse-J and S shaped field lines.  A prominent dip touching tangentially the magnetogram level is displayed in the perspective view in the top right panel of Figure~\ref{Fig_bp} for event 2. A similar topology is evident for event 3 and 4. In the case of event 4, the field line dips are not touching the magnetogram level (z$_{\rm dip} \simeq$ 0.2 Mm above this level) so in this case a BP region is not present at the photospheric level. Still, the inverse S-shape for the field lines is clear. Notedly, these BP/dip locations moves southward in time.  

 As is the case in AR 12371, flux cancellation is a process transforming double J-shaped field lines to an S-shaped FR \citep{ballegooijen1989,Feynman1994} and is associated with the formation of BP with field lines tangential to the magnetogram level. This process is also expected to occur in our studied cases. Converging motions toward PIL drive the flux cancellation reconnection resulting in long FR field line and evidently the field lines of NLFFF along PIL include crossed inverse-J and S.  Finally, we notice that the BP locations   evolve along the PIL before CMEs (not shown) so that magnetic reconnection, building the FR, is expected to occur at intermittent positions along the PIL.

% {\S}{\bf --- General physics expected with BPs } \\
BP separatrices are preferential sites of thin current sheets due to persistent photospheric shearing motions. The reconnection of field lines in these current sheets produces high temperature emission visible in hot AIA passbands. As reconnection progresses dynamically, the FR builds by added flux. In this phase, it has been shown by MHD numerical simulations that the BP bifurcates resulting in the transformation of a BP separatrix to a QSL \citep{aulanier2010}. Then  the standard coronal tether-cutting reconnection sets in below the FR elevating its main body. Later on, the FR becomes unstable leading to the flare onset \citep{aulanier2010,savcheva2012a,savcheva2012c}.  The snapshot of NLFFF model only captures static structure indicating BP topology i.e., the global magnetic topology evolution but not the dynamic reconnection.  

\begin{figure*}[!ht] 
	\centering
	\includegraphics[width=.99\textwidth,clip=]{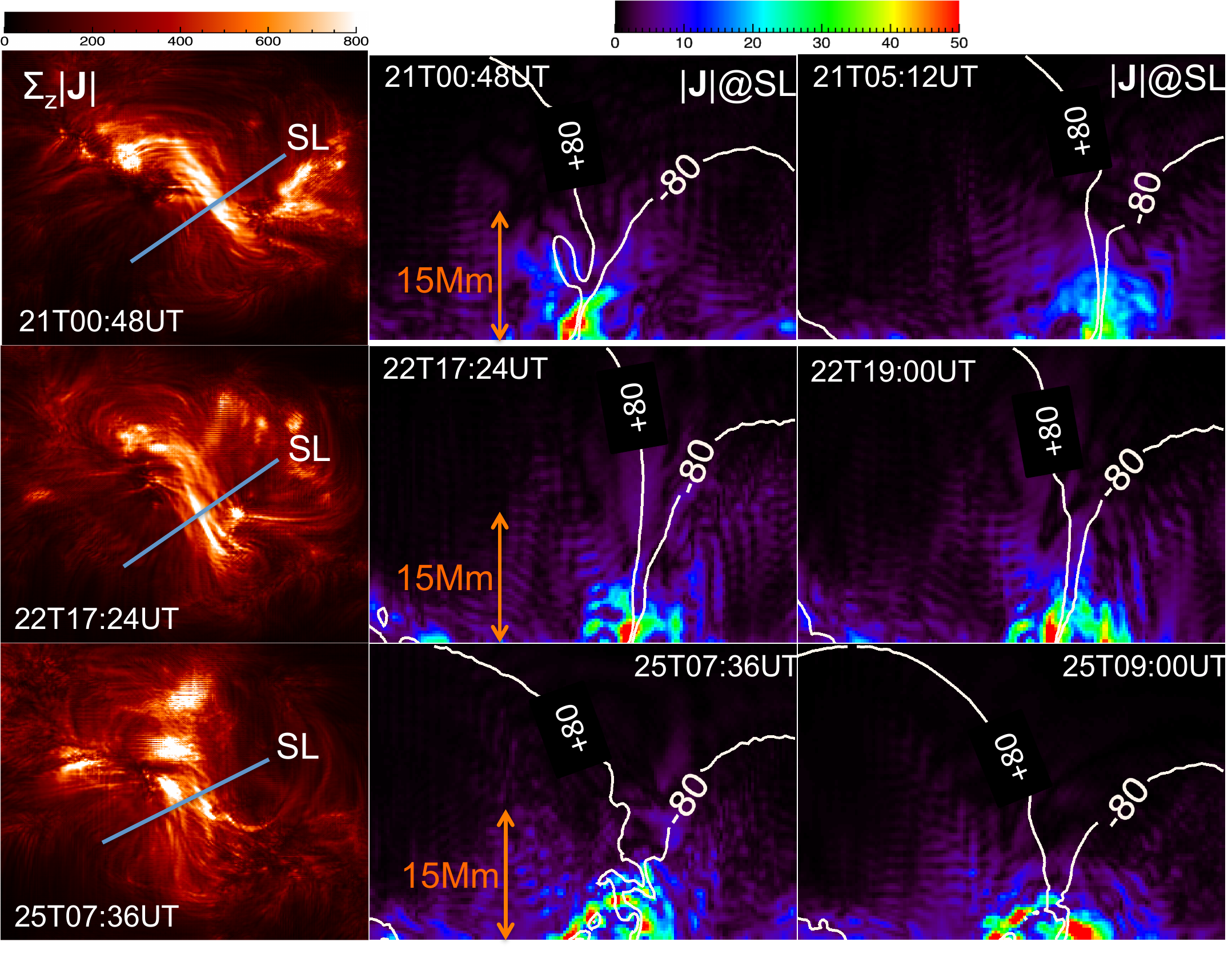}
	\caption{Current distribution just before and after three CME events. 
		\textit{First column:} Vertically integrated maps of $|\mathbf{J}|$ scaled within the interval [0,800]$A\,m^{-2}$ (see the top color bar). They are mimicking the sigmoid observed in EUV (Figure~\ref{Fig2}).  The position of the vertical slice SL, used in the right panels, is shown by a blue line. 
		\textit{Second and third columns:} Electric current density, $|\mathbf{J}|$, in the vertical slice before and after eruption. Contours (white curves) of the vertical component $B_z$ at $\pm 80$ G are over-plotted in each panel. 
		$|\mathbf{J}|$ is scaled within the interval $[0,50]~mA~m^{-2}$ (see top color bar).  	 
		The strong $|\mathbf{J}|$ region corresponds to the sheared core field and it is surrounded by weaker currents present in the arms of the two elbow magnetic structures (Figure~\ref{Fig_core}). 
		The height scale is indicated in the panels of the middle column.
	}
	\label{Fig_curr}
\end{figure*}

%  {\S}{\bf --- Back to studied events! } \\
Moreover, the precursor brightening (4th row panels in Figure~\ref{Fig2}) is co-spatial with the twisted core field (Figure~\ref{Fig_core}). The brightening commences around an hour before the main eruption with the appearance of an hot inverse-S trace in the core, commencing the bright ribbons in the flare phase. Then, the forming FR by flux cancellation detaches the photosphere earlier on, favoring coronal tether-cutting reconnection, as in the numerical simulations reported above.  

\subsection{Coronal Current Distribution}
\label{Coronal-current}
Owing to time evolution of coronal magnetic field by persistent slow photospheric shearing motions, we see a transition of potential-like arcade in the post eruption phase of previous eruption to double inverse J-shaped sheared arcade (this transition is analyzed in Section \ref{Buildup-core}). This process naturally builds coronal volume currents in the stressed configuration. To have more insights on the coronal current distribution, we examine direct volume rendering of $|\mathbf{J}|$ (not shown). Near the magnetic field concentrations, $|\mathbf{J}|$ is stronger since $|\mathbf{J}|$ is proportional to $|\mathbf{B}|$ along a field line in a force-free field. Especially, intense $|\mathbf{J}|$ are present above the inner bipolar region corresponding to the sheared/twisted field structure.

Further, we vertically integrated $|\mathbf{J}|$ in the coronal volume and the resulting 2D maps are shown in first column panels of Figure~\ref{Fig_curr}. The overall morphology of this current distribution is similar to the sigmoid observed in EUV (Figure~\ref{Fig2}). 
For the event 4 of June 25, the $|\mathbf{J}|$ distribution in the core is somewhat less intense than for the two previous events.  This is associated with difference in compactness and twist of the field structure seen in panel (C2) of Figure~\ref{Fig_core}. 

\begin{figure*}[!ht]
	\centering
	\includegraphics[width=.95\textwidth,clip=]{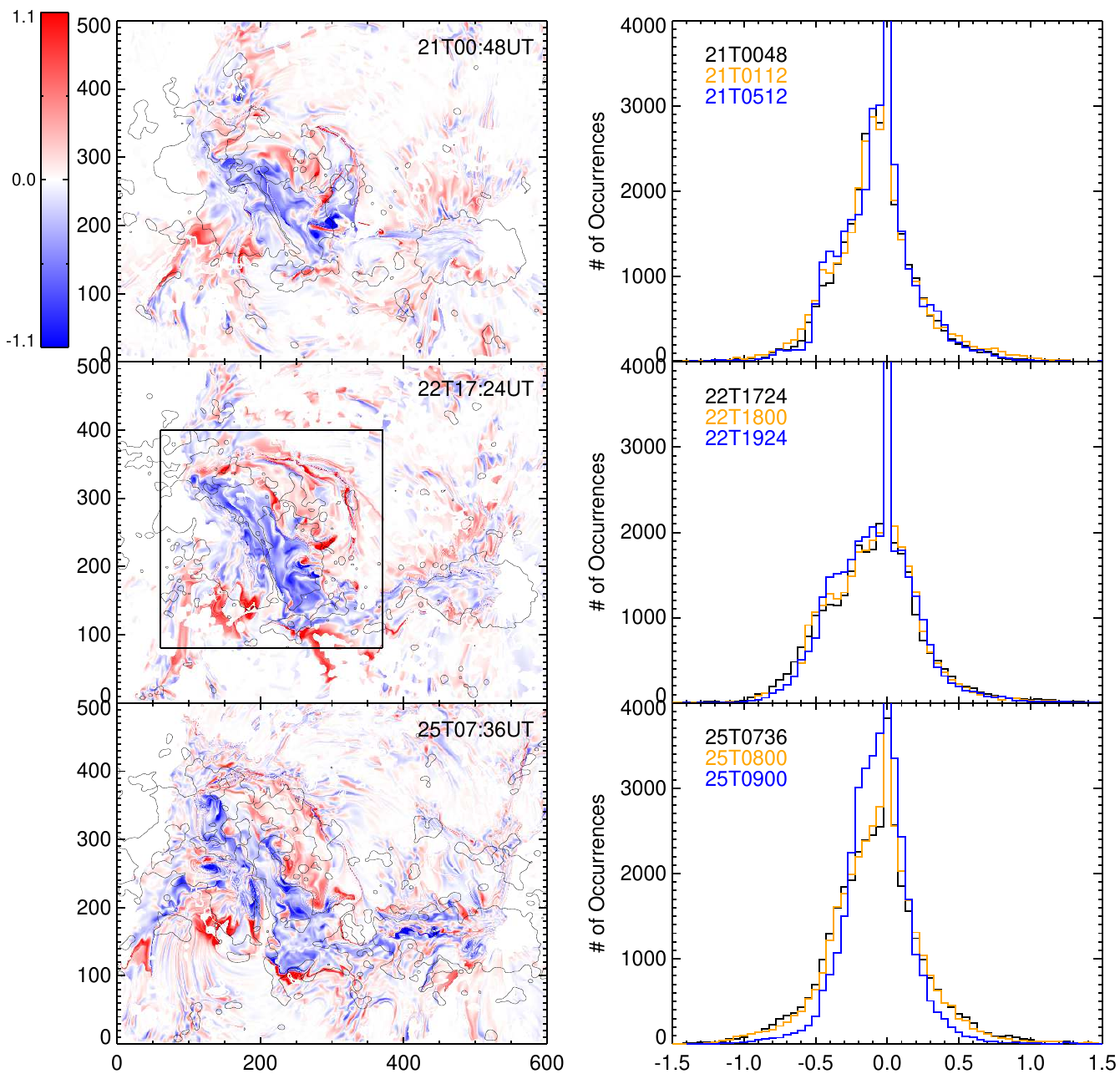}
	\caption{Spatial distributions and histograms of the field line twist (see Equation (\ref{Eq-twist})).
		\textit{Left column:} $\Tw$ distribution of AR 12371 about half an hour before the onset of CME events.  Red (blue) color pixels refer to positive (negative) twist.  Maps are scaled within $\pm 1.1$ turns as shown by the blue-red color scale.  
		Field lines with foot points around the PIL have higher negative twist. They are surrounded by positive twisted field.  A smaller rectangular region, as show in the 22T17:24 UT (second) panel, is chosen for the histogram analysis. 
		\textit{Right column:} Histograms of $\Tw$ before and after the onset of each CME. The asymmetry in the distributions is due to a dominant negative $\Tw$ in the rectangular region. The histograms shrink slightly in width over time owing to relaxation of the field during CME eruption.  
	}
	\label{Fig_tst_dist}
\end{figure*}

Corresponding to before and after the eruption, in the second and third column panels of Figure~\ref{Fig_curr}, we display $|\mathbf{J}|$ obtained in vertical slice planes placed across the sigmoid. The locations of these slices are shown in the left panels of Figure~\ref{Fig_curr}. White curves represent $B_z$ contours ($\pm 80$ G) extracted in the same slice. Intense $|\mathbf{J}|$ corresponds to AR core that is confined below about 15 Mm. The elbow field lines (Figure~\ref{Fig_core}) contribute to the distributed coronal current on either side of the PIL, whereas above the PIL the intense current concentration corresponds to twisted core field. 

The smaller height of coronal current in June 25 panel likely indicates less stressed field owing to less compact sheared core as seen in Figure~\ref{Fig_core} (bottom row). These results delineate that the pre-eruption configuration comprises a sheared/twisted core with intense coronal electric currents above the PIL. For events 2 and 3 (top and middle rows)  the electric current is more concentrated before than after the eruption. After the eruption, in the last column panels, the coronal current is less dense due to expansion of the core field.

\subsection{Field Line Twist Maps}
\label{twist}

%  {\S}{\bf --- Method to Compute Twist Maps} \\
Given 3D field distribution, one can calculate the twist number for each field line \citep{berger2006,inoue2011, ruiliu2016}
\begin{equation} \label{Eq-twist}
\Tw = \int\limits_{L} \frac{\mu_0 \mathbf{J}_{||}}{4\pi B} ~\rmd l
    = \int\limits_{L} \frac{\nabla \times \mathbf{B} \centerdot \mathbf{B}}{4\pi B^2} ~\rmd l
\end{equation} 
Here the twist is related to the parallel electric current given by ${{\mathbf{J}}_{||}}=\frac{\mathbf{J}\centerdot \mathbf{B}}{\left| B \right|}$ and the line integral is along the selected magnetic field line of length $L$. If the magnetic field is force free, $\nabla \times \mathbf{B}=\alpha \mathbf{B}$, then 
$\Tw =\frac{1}{4\pi}\int\limits_{L}{\alpha \,dl} = \frac{\alpha L}{4\pi}$ since $\alpha$ is constant for each individual field line. A $4\pi$ factor in the denominator includes a factor that converts radians into turns and a factor (2) relating local twist (radians per unit length) with $\alpha$ \citep{longcope1998}. In the extrapolation, the magnetic field is not exactly force-free so we rather perform the integration with $\nabla \times \mathbf{B}$ within the integral.   $\mathbf{B}$ and $\nabla \times \mathbf{B}$ are computed at any point in the volume by tri-linear interpolation. Then, the line integral is carried out by a five-point Newton-Cotes formula using the procedure \texttt{int\_tabulated.pro} available in IDL. For a finer structure of $T_w$ distribution, the field line integration is performed on a grid of spatial resolution double that of extrapolation grid. The field lines, that reach one lateral or top boundary, are assumed to be open, and they are set with zero value of $\Tw$.  

%  {\S}{\bf --- Results of twist maps} \\
The resultant $\Tw$ maps at the times before the onset of CMEs are displayed in left column panels of Figure~\ref{Fig_tst_dist}. We find that the field line twist magnitude does not exceeds 1.1 turns across the maps at different times.  Therefore, we scale the maps within $\pm 1.1$ turns on a blue-red color map. The central core has always negative values of $\Tw$ ranging from -1.0 to -0.4 turns. In all the cases, a highly-twisted core, surrounded by $\Tw$ values with opposite sign, is observed.  

%  {\S}{\bf --- Histograms} \\
For a more quantitative analysis, we choose a smaller rectangular region surrounding the core of the sigmoid for the three events (its extension is shown in the central left panel of Figure~\ref{Fig_tst_dist}), and perform an histogram analysis of $\Tw$ at pixels with $|B_z|>120$G. Within this region, the fraction of pixels having more than one turn is  1.4\%, 1.2\%, 1.46\% respectively before these successive CMEs. Similarly, the fraction of pixels with $0.5<|\Tw|<1$ turn is still small since it is limited to 11.8\%, 15\%, 12\% respectively of the total number of pixels with $|T_w|>0$ within the selected rectangular region.

 % {\S}{\bf --- Compare to other studies} \\
We compare the above results with the ones obtained for another AR where twisting motions were analyzed \citep[AR 10930][]{inoue2011}. The fraction of pixels in the above two intervals of twist is 5\% and 40\%, respectively.  Therefore, it appears that rotational flux motions are more efficient in injecting twist and helicity than shear and converging motions, as expected.  Further, according to \citet{torok2004}, the critical twist to destabilize the ideal MHD kink modes in an anchored magnetic loop is about 1.75. Therefore, the observed low twist distribution in the studied AR suggests that the kink instability is not a relevant process to trigger these eruptions. 

 % {\S}{\bf --- Evolution of the distribution} \\
 The right column panels of Figure~\ref{Fig_tst_dist} show histograms of $T_w$, where each panel shows variation of $T_w$ distribution during the CME eruption. To show how $T_w$ vary, we compute $T_w$ at two epochs before eruption and one epoch after eruption and plotted in different colors (blue color histogram is after eruption). Note that the asymmetric distribution is a signature of dominant negative twist in rectangular region corresponding to core flux. In all cases, the occurrences in the histogram wings is lower than that before the CME. To quantify this decrease of twist owing to magnetic energy release, we count the pixels with $|T_w|$ above and below 0.3 turns and express in terms of percentage change from higher values ($>0.3$) to lower values during pre-to-post eruption evolution. At the epochs before CME eruption, the percentage of pixels above 0.3 turns is 28\%, 31\%, 32\% respectively, suggesting increasing twisted flux from first to last event. For the three CME cases on 21, 22 and 25 June, we found that the above percent of pixels decrease to 26\%, 29\%, 16\% (so a decrease of  2\%, 2\%, 16\%) respectively. The corresponding effect is seen with the histograms, where the histogram wings in the post eruption shrink below that for pre-eruption. 

\begin{figure*}[!ht]
	\centering
	\includegraphics[width=.99\textwidth,clip=]{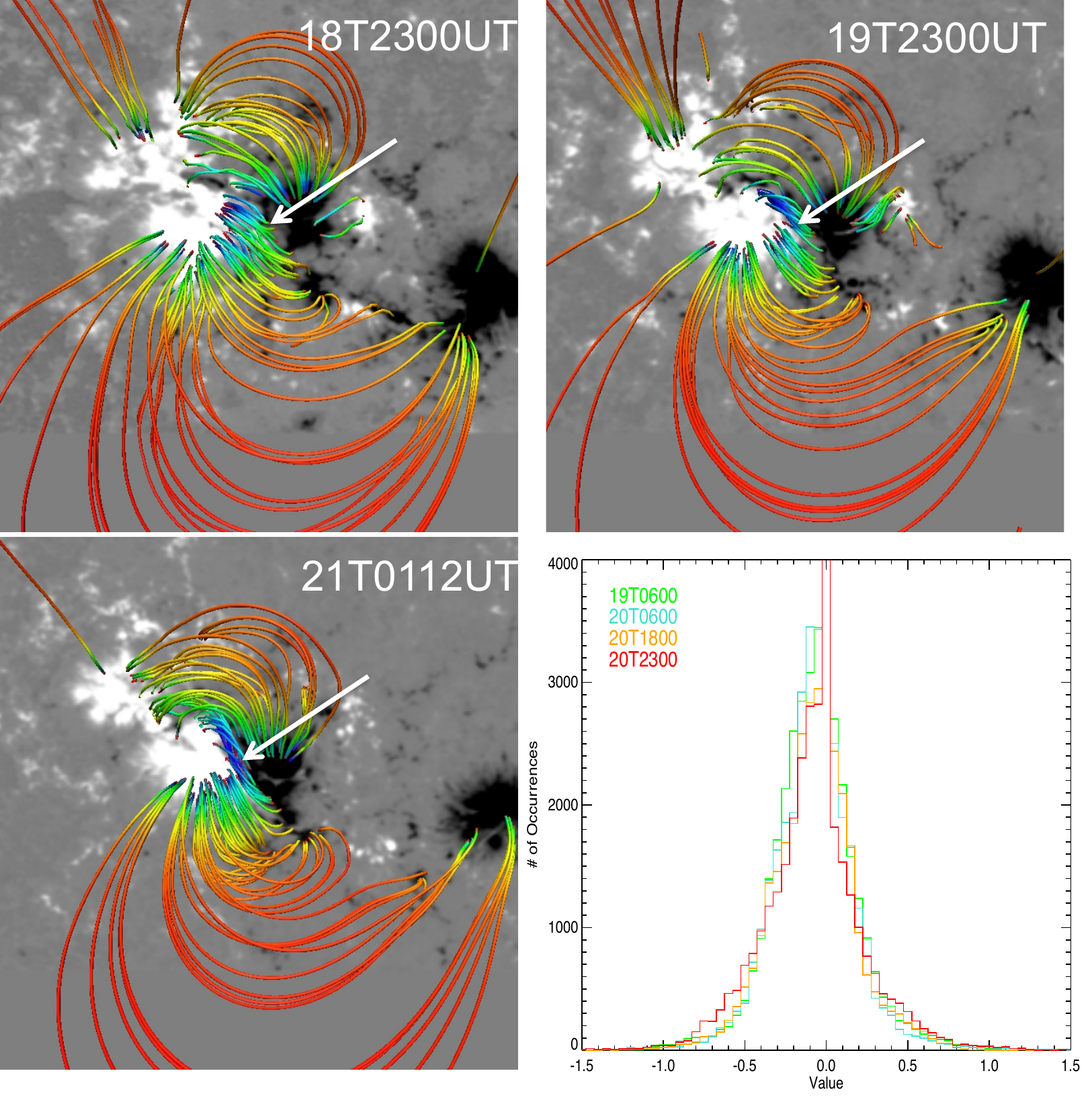}
	\caption{Development of twisted core field by shearing motions over two days of quasi-static evolution after CME1. 
		\textit{Image panels:} NLFFF field lines are over-plotted on magnetograms in the post eruption evolution of first CME (starting at 17:24UT on June 18) and before the second CME (starting at 02:30UT on June 21).  The first time is just after CME1 and the next ones are separated by about one day.  Arrows point to the buildup of the twisted core field. Field lines are color coded (blue (red): $\sim$1200 (2)G) with the horizontal field strength in height. 
		\textit{Bottom right panel:} Histograms of $\Tw$ (Equation (\ref{Eq-twist})) at four different times.  It shows increasing wings corresponding to the development of a twisted core field. }
	\label{Fig_tst_dev}
\end{figure*}

\subsection{Buildup of a Sheared/Twisted Core Field}
\label{Buildup-core}
As AR magnetic structure evolves, the formation of a twisted core field from an initial arcade configuration is expected under the effect of shearing and converging motions. To show this, the NLFFF reconstruction is applied at different times in the post eruption evolution of CME1 on June 18. In Figure~\ref{Fig_tst_dev}, we show magnetic structure of the AR at three instances. The first snapshot is in the relaxing phase present after the CME1.  The core field has a remnant shear besides elbow field lines. In the course of further evolution driven by slow shearing/converging motions, the core field becomes twisted gradually. In particular, in the later two snapshots (pointed by white color arrow), the core field appears more like a bundle of twisted field lines (FR). In addition, the elbow field lines appear more compact than earlier on. This buildup of twisted core is a key process of energy storage in the sigmoid structure which occurs to reform the FR from the pre-eruption arcade.  This leads to the launch of the CME2 at 02:00\,UT on June 21.

This development of a twisted core is further studied quantitatively by utilizing the twist $\Tw$ maps derived from the NLFFF extrapolations. Similar to earlier section, in the last panel of Figure~\ref{Fig_tst_dev} we display histograms of $\Tw$ maps at four times in the aftermath evolution of the CME1. This time evolution of histograms delineate the transfer of pixels about the center (with $|\Tw|<0.5$) toward the wings ($|\Tw|>0.5$). The fraction of pixels with $|\Tw|>1$ turn shows a slight increase from 0.4\% to 1.2\% from first instance to the last. However, corresponding to the buildup of the sheared core, the fraction of pixels with $0.5<|\Tw|<1$ is significant and growing from 6.1\% to 11.3\%,  so almost by a factor 2. If we rather select the interval $0.3<|\Tw|<1$, the fractional increase is lower as it grows from  21.2\% to 29.7\% from first instance to the last one, so increasing only by a factor $\approx 1.4$. 

\begin{table}[!ht]
	\centering

		\caption{Magnetic energy in units of $10^{32}$ ergs before and after each CME eruption in AR 12371. $\Ep$, $\Et$ and $\Ef$ are respectively the potential, total, and free magnetic energy ($\Ef=\Et-\Ep$).}
		\begin{tabular}{cccc}
			\hline 
			event (time) & $\Ep$ & $\Et$ & $\Ef$ \\ 
%			event (time) & $\Ep$ ($\times 10^{32}$ ergs) & $E_T$ ($\times 10^{32}$ ergs)  & $\Ef$($\times 10^{32}$ ergs) \\ 
			\hline 
%			CME1(2015-06-18T16:12\,UT)	&30.48	&32.24	&1.75\\
%			CME1(2015-06-18T18:00\,UT)	&30.95	&32.69	&1.46\\
			CME2(2015-06-21T00:48\,UT)	&26.25	&28.39	&2.14\\
			CME2(2015-06-21T04:00\,UT)	&25.97	&28.64	&2.09\\
			CME3(2015-06-22T17:24\,UT)	&21.54	&24.19	&2.64\\
			CME3(2015-06-22T19:24\,UT)	&21.75	&24.23	&2.47\\
			CME4(2015-06-25T07:36\,UT)	&22.58	&24.03	&1.44\\
			CME4(2015-06-25T09:00\,UT)	&22.81	&23.80	&0.98\\
			\hline 
		\end{tabular} 
		\label{tab1}

\end{table}

\subsection{Magnetic Energy}
\label{energy}
Magnetic free energy ($\Ef$) is a measure for excess energy available for the eruptive flares. It is estimated by subtracting the potential field energy ($\Ep$) from total magnetic field energy ($\Et$). To study the height variation of the magnetic free energy, we compute the surface integral of free energy as (e.g., \citealt{mackay2011,vemareddy2016b})
  \begin{equation}  \label{Eq-energy}
  \Efs (z) = \int\limits_S \frac{B^2}  {2 \mu_0} ~\rmd x ~\rmd y
          -\int\limits_S \frac{\Bp^2}{2 \mu_0} ~\rmd x ~\rmd y
  \end{equation}
from the NLFFF and potential fields of the AR. For this purpose, we consider the time frames before and after the eruption. Specifically, we choose the pre-flare snapshot just before the first observed brightenings and the post eruption snapshot at the end of the flare decay phase ensuring the field relaxation gets captured in the vector magnetograms.

We caution that our computations of free energy estimation is subjected to errors due to projection effects, extrapolation model, and noise/bias in input observations. Moreover, implicit smoothing in preprocessing procedure also contributes underestimating the actual coronal free energy content.

\begin{figure}[!ht]
	\centering
	\includegraphics[width=.48\textwidth,clip=]{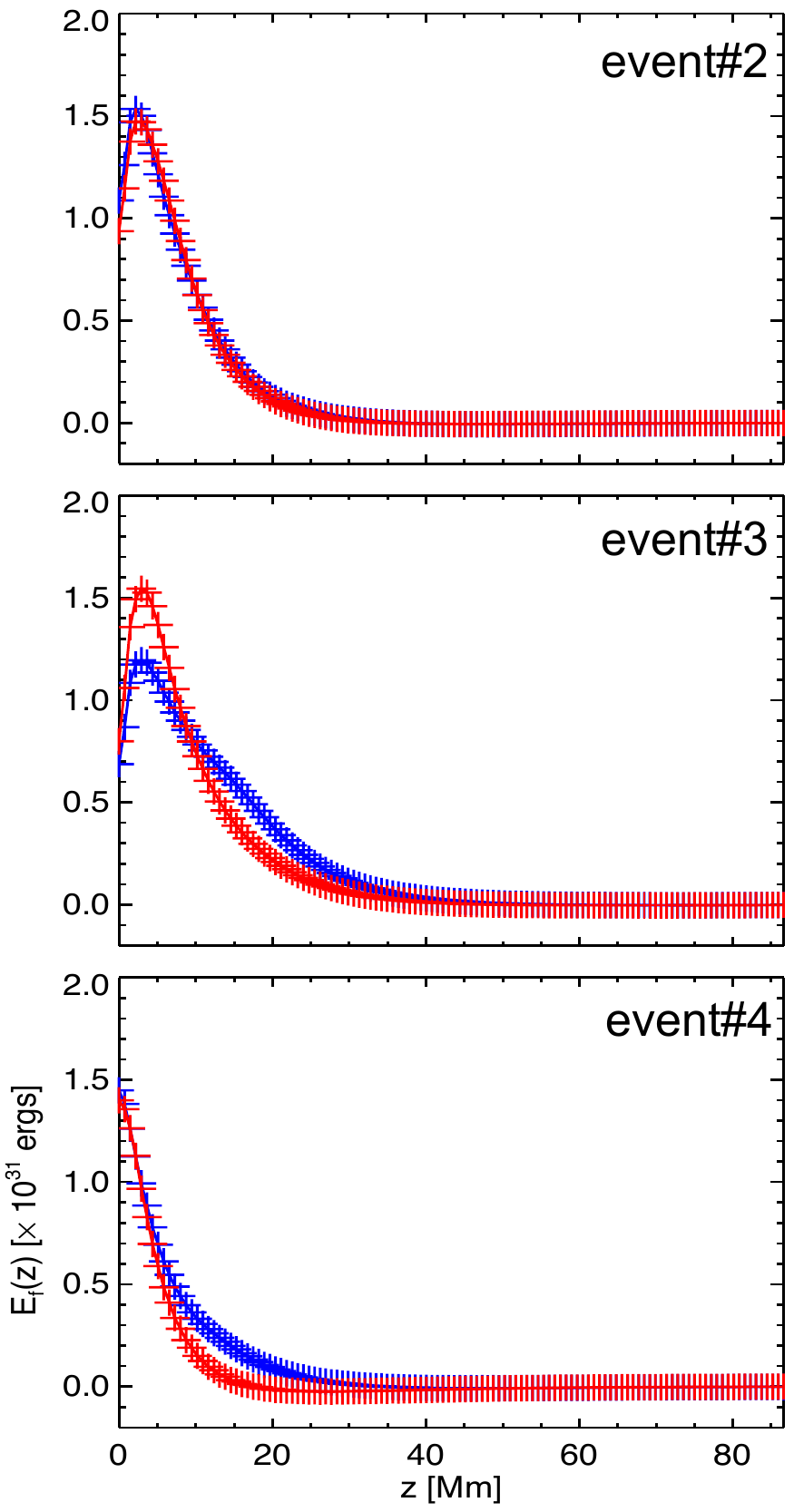}
	\caption{Free energy $\Efs$, Equation (\ref{Eq-energy}), in horizontal planes as a function of height (z, in Mm).  Blue (red) curve represents $\Efs$ before (after) eruption (Table~\ref{tab1}). Most of the free energy is located below 20 Mm. On average, the post eruption curves are below pre-eruption curves, indicating that, as expected, $\Efs$ drops through the CME eruption. }
	\label{Fig_ene_arr}
\end{figure}

During the three CME cases the plots in Figure~\ref{Fig_ene_arr} delineate that most of the coronal free energy is contained below 20 Mm and $\Efs$ is maximum a low height, below 5 Mm. This energy is related to the coronal electric current distribution depicted in Figure~\ref{Fig_curr}. 
$\Efs$ decrease for event 2 is small, while for events 3 and 4 the decrease of $\Efs$ is significant in the range 10-30 Mm. Only in event 3 an increase of $\Efs$ is present at low height.

In Table~\ref{tab1}, the global energies for all three CME events are reported. In each case the time is considered just before and after the flares as above. $\Ep$ and $\Et$ are of the order of $10^{33}$ erg, which is the typical energy content of large ARs (e.g., AR12192 \citealt{jiangc2016b}).  From event 2 through 3, $\Ep$ drops gradually. 
This is consistent with the photospheric magnetic flux which is decreasing both by dispersion and by cancellation at the PIL. A decreasing $\Ep$ with reducing AR flux content is also observed in AR 11283 in the study by \citet{jiangc2014}. 
In all events the potential energy is almost the same before and after (Table~\ref{tab1}) as expected because of the small observed evolution during the events of the vertical field component at the photospheric level. The magnetic energy available for an event, which is converted e.g. as kinetic, thermal and radiative energies, is mostly the difference of coronal magnetic energy between after and before the event since eruptions are involving the ejection of closed magnetic structures and since the energy input by photospheric shearing motion is small compare to the energy accumulated between two events (with a constant flux rate, this ratio is the ratio of the durations, so of the order of $2/24 \approx 0.1$).

The magnetic free energy drop for the three events is $0.05$, $0.17$, $0.46$ in units of $10^{32}$ ergs, respectively, accounting for 2.3\%, 6.4\%, 32\% of the pre-eruption free energy. These small values are qualitatively consistent with the estimates of twist number variation in Section \ref{twist}, where only a small number of field lines (upto 16\%) show a variation of twist number corresponding to field reconfiguration. 

In summary, our analysis suggests that magnetic energy is being pre-stored in the magnetic structure by gradual shearing motions around the PIL, then it is released by intermittent eruptions. 

\begin{figure}[!ht]
	\centering
	\includegraphics[width=.48\textwidth,clip=]{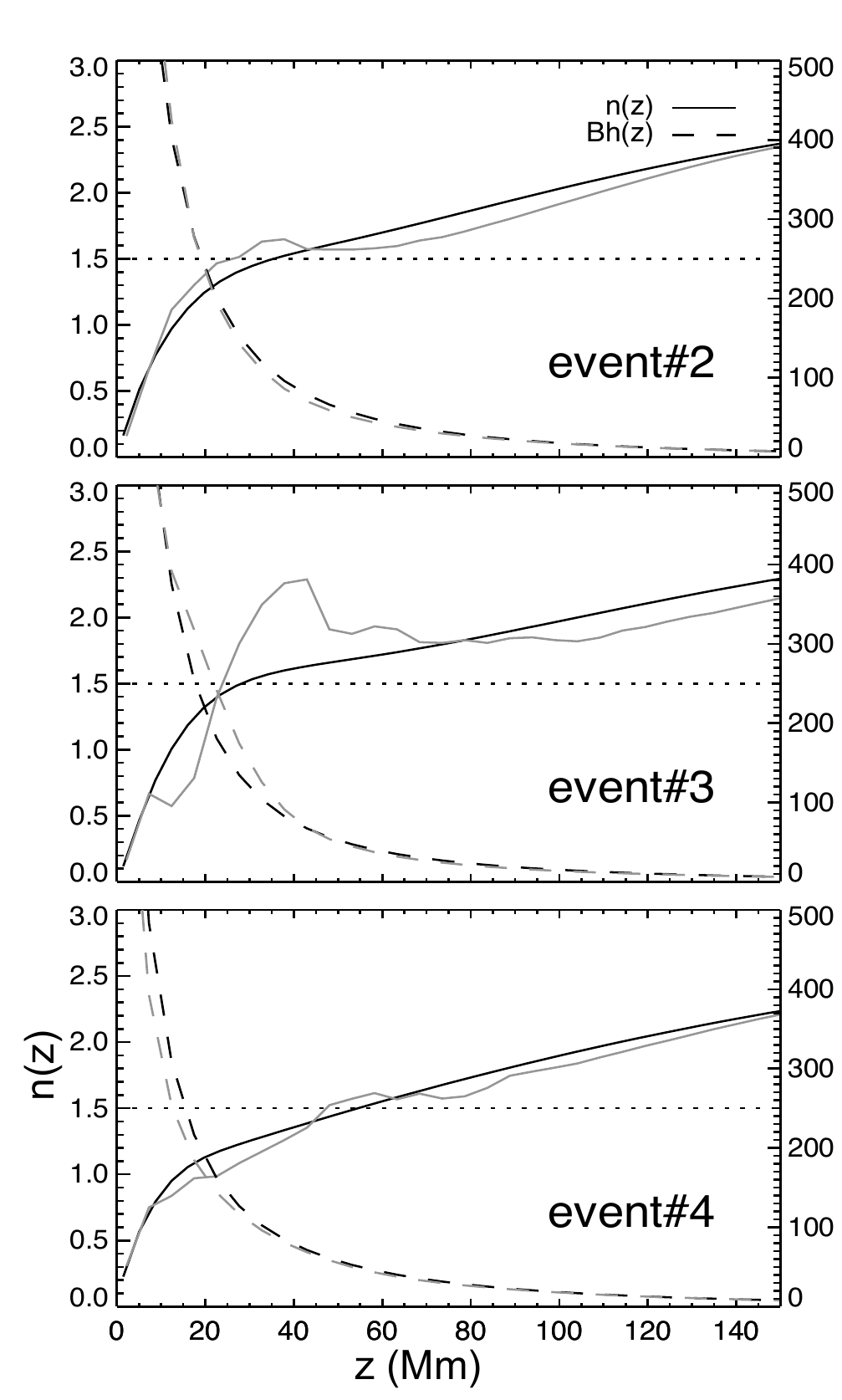}
	\caption{Decay index (n) of the background field (ordinate) as a function of vertical height (abscissa, in Mm). Black (gray) continuous curve represents the mean decay index obtained from the potential (NLFFF) field, respectively. The horizontal field strength ($B_h$ [G]) is also plotted at the same spatial locations with dashed lines within a range between 0 and 500 G (see the scale located on the right side of panels).  Dotted horizontal line indicates the critical value of $n_{crit}=1.5$. For all three events, this critical value of $n$ corresponds to a height just below  or $\approx 40$ Mm.
	}
	\label{Fig_dc_ind}
\end{figure}
\subsection{Torus Instability}
\label{Torus-Instability}

To reveal the role of background field, we also compute the decay index $n(z)=-d\log ({{B}_{h}})/d\log z$, where $B_h$ is the horizontal component of the background field and $z$ is the height above the photosphere. Since the vertical magnetic-field component does not contributes to the inward confining force, the decay index is computed only with the horizontal components of the field. This background field strength decides the torus instability (TI) criteria. \citet{torok2005} proposed the value of 1.5 as a critical decay index.  In fact, the critical value of $n$ depends on a number of factors, but is expected to be in the range $1.1\le n_{crit}\le 2$ \citep{bateman1978,kliem2006,demoulin2010}. 

In the present study, $B_h (z)$ is obtained at different spatial points located along the PIL in the inner bipolar region. Then an average value of $n(z)$ is derived corresponding to $B_h(z)$ at those points in both potential field and NLFFF (grey and black curves in Figure~\ref{Fig_dc_ind}). 

The theory separates the FR magnetic field from the background field and $n(z)$ is only computed from the background field. This separation is trivial because analyzed analytical models are constructed by the superposition of these two fields. However, with magnetic extrapolations there is no precise way to separate the FR field, or its associated electric currents, from the background. Then, we consider two extreme limits: all the currents are associated with the FR, so the background field is potential, and at the opposite the FR has a negligible fraction of electric currents, so the background field is the full NLFFF. These are clearly very extreme cases but we show below that the results are comparable so that we do not need to define what is the field associated to the FR for this study of the decay index.

In Figure~\ref{Fig_dc_ind}, we plot $n(z)$ above z=0 plane. $B_h$ is also plotted (in units of Gauss) with its scale reported on the right of each panel. In all three cases, the $n(z)$ for potential field and NLFFF follows each other except short scale spikes in NLFFF arising due to differentiation of a more structured field. For all three cases, the characteristic curve $n(z)$ approaches the critical value, set to $n_{crit}=1.5$, at a height between 20 and 50 Mm, indicating a favorable magnetic environment for the destabilization of the FR, then the launch of a CME when the magnetic evolution brings the FR axis to such height. There is some height difference for $n_{crit}$ (up to 10 Mm) for NLFFF and potential field, but the average critical height is around 40 Mm. 

The TI domain in AR 12371 is comparable to other eruptive ARs, for instance, the flare-CME productive ARs 11158, 11429  have TI critical height of 42, 34 Mm respectively \citep{sunx2015}. Notedly, the horizontal field at 42~Mm in these eruptive cases is ($\sim50$~G), so smaller by a factor of four than in confined AR 12192 (230~G).  ARs with failed eruptions have a distinct TI height range compared to ARs with successful eruptions. Recent observations found that the critical height for TI can be as high as 236~Mm \citep{wangw2017}.  For a confined eruption studied by \citep{guoy2010}, $n_{crit}$ is always above 100~Mm suggesting a stronger restraining field over the FR. Similarly, the CME less AR 12192 is found to have stronger overlying field with a TI critical height up to 77~Mm \citep{sunx2015}. Indeed, the super AR 12192, producing large flares without CMEs, is found to have more closed field flux above than in the sheared core field surrounding the PIL \citep{sunx2015,jiangc2016b}. 

\section{Summary and Discussion}
\label{summ}
% general evolution scenario and theoretical expectations; connecting paper I
We study the magnetic structure of AR 12371, by modeling the coronal field with the photospheric vector magnetic field measurements, in relation to the erupting sigmoid observed successively over days. 
The AR coronal magnetic field is mainly driven by continued shear motions at the photospheric level, which inject helicity flux of dominant negative sign (See Figure 8 in \citealt{vemareddy2017b}). Being co-spatial with the sigmoid, the injected helicity flux is mostly included in the sheared/twisted field of the sigmoid. As the coronal field can not accommodate indefinite an accumulation of magnetic helicity, a CME is inevitable \citep{low1994}, then the stored helicity is partly ejected with the FR.

% modelled strcture supporting tether-cutting reconnection, no BP explanation
In the present study of the three events, the model field structure of pre-eruptive sigmoid has a low-lying twisted core and an overlying arcade resembling the sigmoid morphology observed in coronal EUV images. The NLFFF core is dominantly embedded in a large scale sheared arcade with the inner part along the PIL having opposite J-shaped field lines with crossed legs. Closer and above the PIL inverse-S field lines with dips touching tangentially the photosphere are present (Figure~\ref{Fig_bp}). This kind of topology is a manifestation of BP-separatrix field lines formed by photospheric reconnection of opposite-J field lines and indicates the formation of weakly twisted FR at the inner core of the sigmoid embedded in the large scale sheared arcade \citep{antiochos1999}. As shown in previous simulations (see Section~\ref{Intro}), a larger FR is progressively build up by reconnection at the photospheric level, then in the corona below the FR.  At some point of the evolution the FR become unstable which leads to a CME.  

% Signatures of quasi-static evolution inferred from NLFFF
Moreover, the NLFFF modeled structure captured major features of energy storage and release mechanism, viz., sigmoid-to-arcade-to-sigmoid transformation, that is being recurrent under continuous photospheric flux motions. Calculations of the field line twist reveal an increase of pixels  ($\approx$ 7\% averaging over the events) having a range of field line twist ($0.3<T_w<1$), indicating the quasi-static buildup process of twisted core field by slow shearing and converging motions around the PIL. Similarly, we observed a decrease of pixels within the same range of twist, consistent with the decrease of free energy corresponding to the field reconfiguration from sigmoid to potential like arcade during the eruption.  

The magnetic evolution in this AR is in contrast to the confined AR 12192 which has a larger helicity flux but a much weaker (by a factor of 10) normalized coronal helicity content ($H/\Phi^2$). As inferred by \citet{vemareddy2017b}, a small value of $H/\Phi^2$ implies a large flux content unrelated to sheared/twisted part. This flux is present as an overlying flux to stabilize the FR.  This is complemented by the results of $n(z)$ in the AR field. The critical height in eruptive AR 12371 is around 40 Mm while in confined AR 12192, it is at larger height (77 Mm, \citealt{sunx2015}). 

% some remarks 
The ejective nature of AR 12371 differs in some global magnetic properties with respect to those of confined eruptive ARs. The super AR 12192 is an interesting example being not producing any CME even in association with X-class flares and therefore is a non-eruptive counterpart of AR 12371. Data driven modeling of AR 12192, performed by \citet{jiangc2016b}, have shown that the AR field structure remained in the sheared arcade configuration without forming two-J shape like and escaping FR unlike the AR case studied here.

% further work to be done for more insights
Our analysis of observations are based on static modeling of NLFFF of AR magnetic structure. More insights on dynamic aspects of FR formation, its lift off can be gained from data driven MHD modeling of the AR (e.g., \citealt{jiangc2012b,jiangc2016a}) and would be worth of further investigation.

\acknowledgements SDO is a mission of NASA's Living With a Star Program. This work utilizes data obtained by the Global Oscillation Network Group (GONG) Program, managed by the National Solar Observatory, which is operated by AURA, Inc., under a cooperative agreement with the National Science Foundation. We are grateful to the referee's detailed constructive comments and suggestions which improved the scientific presentation of the results. P.V. is supported by an INSPIRE grant of AORC scheme under the Department of Science and Technology.  The NLFFF code is developed by Dr. T. Wiegelmann of Max Planck Institute for Solar System. We thank Jun Chen and Rui Liu of University of Science and Technology of China for a useful discussion on field line twist computation. 3D rendering is due to VAPOR (\url{www.vapor.ucar.edu}) software. We acknowledge an extensive usage of the multi-node, multi-processor high performance computing facility at Indian Institute of Astrophysics. 

\bibliographystyle{apj}
%\bibliography{../../../ref_lib}
%\bibliography{ref_lib}

% FIGURES SECTION %
%%%%%%%%%%%%%%%%%%%%%%%%

\end{document}